\documentclass[runningheads]{llncs}
\usepackage{graphicx}
\usepackage{siunitx}
\begin{document}

\title{MixNet: Multi-modality Mix Network for Brain Segmentation}
%
%
\author{Long Chen\inst{1} \orcidID{0000-0002-5280-4727} \and
Dorit Merhof\inst{1}\orcidID{0000-0002-1672-2185} }
\authorrunning{L. Chen et al.}
%
\institute{
Institute of Imaging \& Computer Vision, RWTH Aachen University, Aachen, Germany\\
\email{\{long.chen,dorit.merhof\}@lfb.rwth-aachen.de}\\
\url{https://www.lfb.rwth-aachen.de/}}

\maketitle              
\begin{abstract}
Automated brain structure segmentation is important to many clinical quantitative analysis and diagnoses. In this work, we introduce MixNet, a 2D semantic-wise deep convolutional neural network to segment brain structure in multi-modality MRI images. The network is composed of our modified deep residual learning units. In the unit, we replace the traditional convolution layer with the dilated convolutional layer, which avoids the use of pooling layers and deconvolutional layers, reducing the number of network parameters. Final predictions are made by aggregating information from multiple scales and modalities. A pyramid pooling module is used to capture spatial information of the anatomical structures at the output end. In addition, we test three architectures (MixNetv1, MixNetv2 and MixNetv3) which fuse the modalities differently to see the effect on the results. Our network achieves the state-of-the-art performance. MixNetv2 was submitted to the MRBrainS challenge at MICCAI 2018 and won the 3rd place in the 3-label task. On the MRBrainS2018 dataset, which includes subjects with a variety of pathologies, the overall DSC (Dice Coefficient) of 84.7\% (gray matter), 87.3\% (white matter) and 83.4\% (cerebrospinal fluid) were obtained with only 7 subjects as training data.

\end{abstract}

\section{Introduction}
\label{sec:intro}

Accurate automated segmentation of brain structures, e.g., white matter (WM), gray matter(GM), and the cerebrospinal fluid (CSF) forms the basis for high-throughput quantitative analyses and associated diagnoses. while computed tomography (CT) and positron emission tomography (PET) is also used for brain structure analysis, magnetic resonance imaging (MRI) is the most popular choice \cite{review1}. We will only talk about MRI in this work. 

As the deep learning approaches are becoming mature, they gradually outperforms previous methods \cite{freeSurfer1,freeSurfer2,fsl,spm}. Based on the network architecture, these deep learning approaches can be roughly divided into two categories: the patch-wise \cite{p1,p2,p3} and semantic-wise \cite{s1} architecture. The patch-wise approach takes a local patch around a pixel as input. Most of the current works use this strategy, because of its efficiency of using the training dataset. Compared to the semantic-wise approach, the patch-wise approach can extract large number of patches from the MRI subjects for training. But unlike unstructured segmentation, brain structures preserve same relative positions in all subjects and patch-wise approaches ignores that information. Some works like \cite{p3} make up for this by augmenting the network input with coordinates of voxels, but semantic-wise methods still have advantages in nature.    

In addition to the overall architecture, we can also use input dimensions to distinguish between different methods. The 3D networks leverage the spatial information more efficiently than 2D networks by sharing kernels across three dimensions. The cost is longer runtime and limited network size. As discussed in Section \ref{subsec:direction}, the 2D network can observe the 3D MRI volume from different directions, that is, more 2D slices as training data. This strategy does not only provides more training images but also plays the role of a ensemble model. By fusing the results obtained from 2D slices along different orientations the segmentation should  be more robust and spatially consistent as well.   

We propose a 2D semantic-wise CNN to handle the brain structure segmentation problem in Section \ref{sec:method}. Three structures are tested to see the effect of different ways of mixing multiple modalities. We call them MixNetv1, MixNetv2 and MixNetv3 in Section \ref{subsec:orga}. The experiments are performed with the MICCAI challenge MRBrainS2018 dataset. The dataset contains annotated multi-sequence (T1-weighted, T1-weighted inversion recovery and T2-FLAIR) scans of 30 subjects. Seven of them are distributed as training data, while the rest subjects are kept unreleased for test. For a limited training dataset, the transfer learning \cite{transfer} usually boosts the overall segmentation results. But this is achieved by using extra data implicitly. Our experiment works with only 7 subjects of the MRBrainS2018 training dataset.

The code developed for this work and trained models will be available online: \url{\textbf{https://github.com/looooongChen/MRBrainS-Brain-Segmentation}}

\section{Method}
\label{sec:method}

In Section \ref{subsec:unit} we introduce the residual dilated convolution unit. Except the initial convolution layer and the output module, MixNet is composed of residual dilated convolution units connected in series or parallel. Section \ref{subsec:orga} discusses different ways of using multi-modalities. Section \ref{subsec:direction} describes the method of acquiring more 2D training slices from the 3D MRI volume. 

\subsection{Basic Units of the Nets}
\label{subsec:unit}

As shown in Figure \ref{fig:v1}, Figure \ref{fig:v2} and Figure \ref{fig:v3}, the networks are composed of three types of basic units: the InitUnit (Figure \ref{fig:input}), the DilateResUnit (Figure \ref{fig:unit}) and the OutputUnit (Figure \ref{fig:output}). In this section, we will described them in detail.

\subsubsection{Initial Unit (InitUnit)}
\label{subsubsec:initUnit}

\begin{figure}
	\begin{minipage}{\textwidth}
		\centering
		\includegraphics[width=0.9\linewidth]{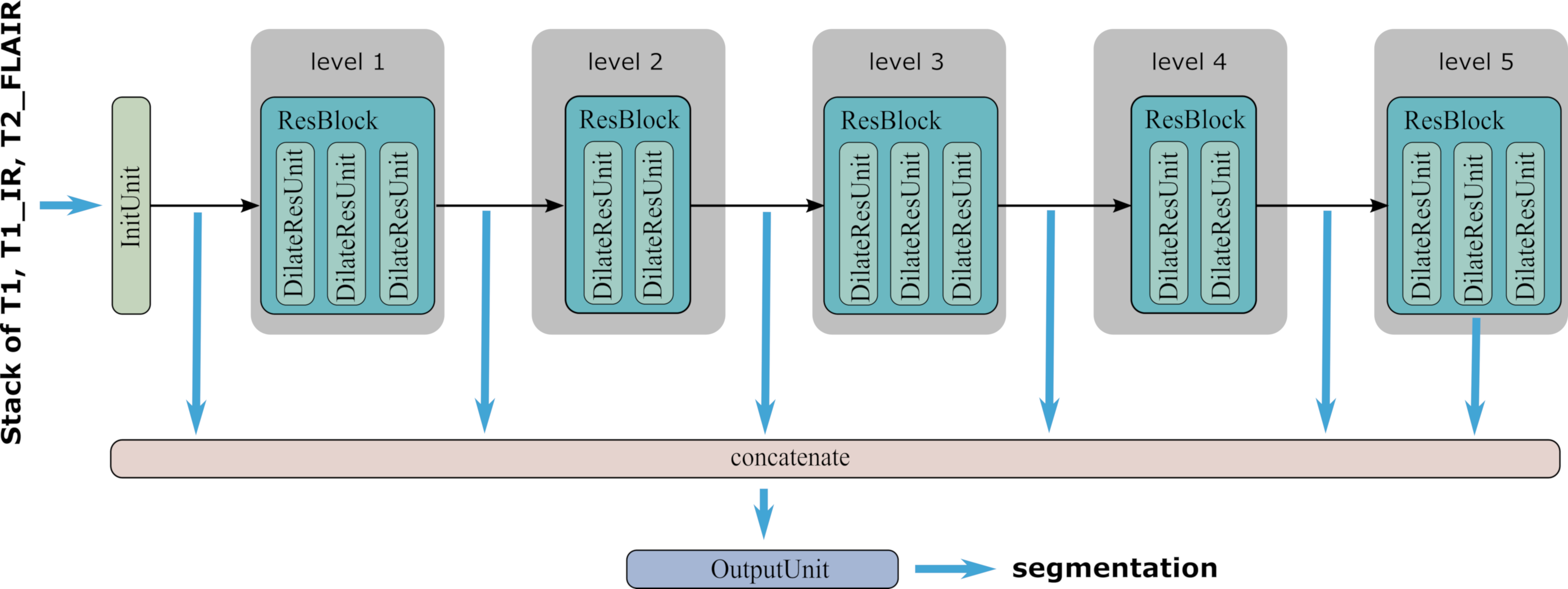}
		\caption{MixNetv1: multiple modalities are stacked at the very beginning.}
		\label{fig:v1}
	\end{minipage}
	\begin{minipage}{\textwidth}
		\centering
		\includegraphics[width=0.9\linewidth]{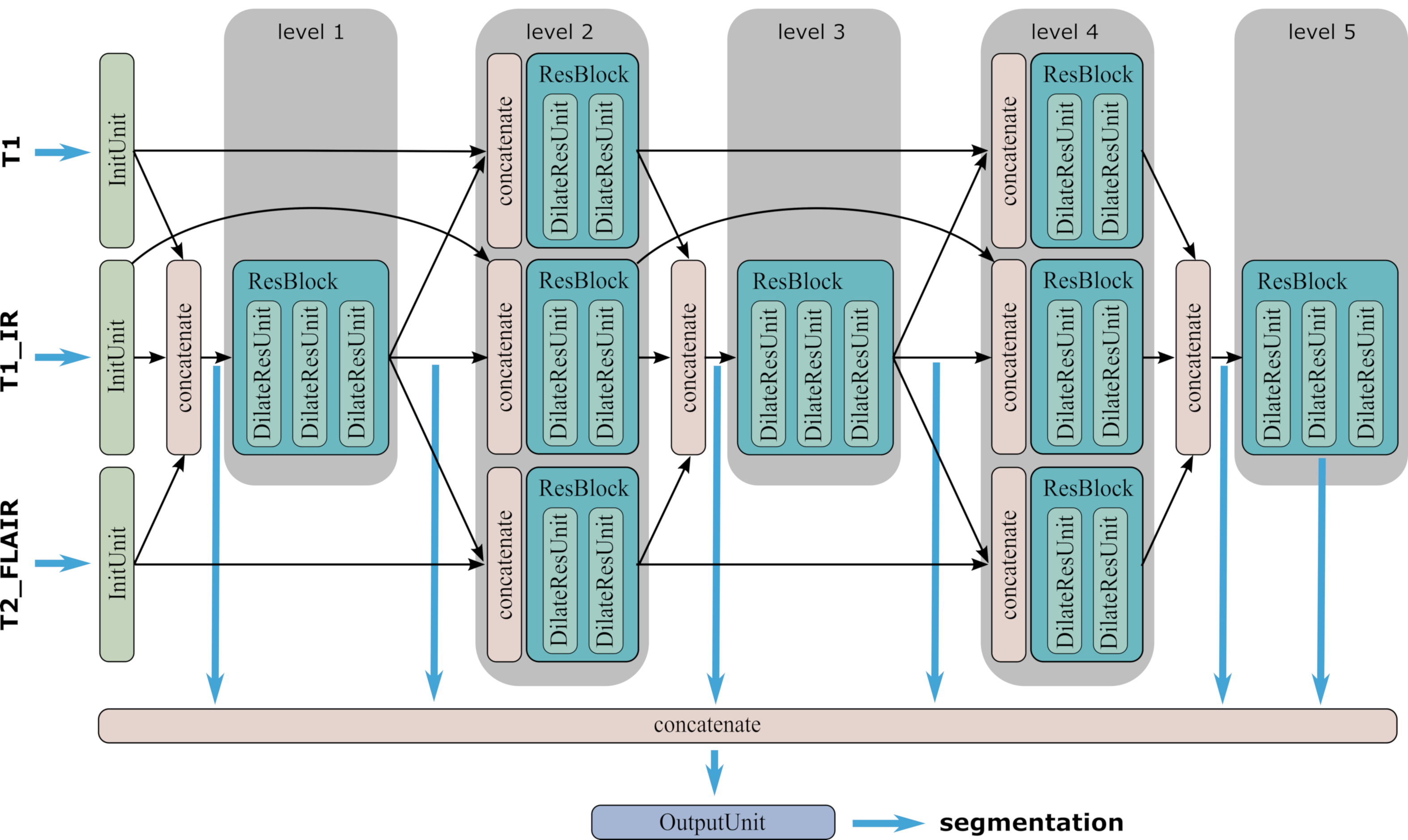}
		\caption{MixNetv2: summarization of multi-modality information is performed periodically, then the summarization is fed back to each modality stream.}
		\label{fig:v2}
	\end{minipage}
	\begin{minipage}{\textwidth}
		\centering
		\includegraphics[width=0.9\linewidth]{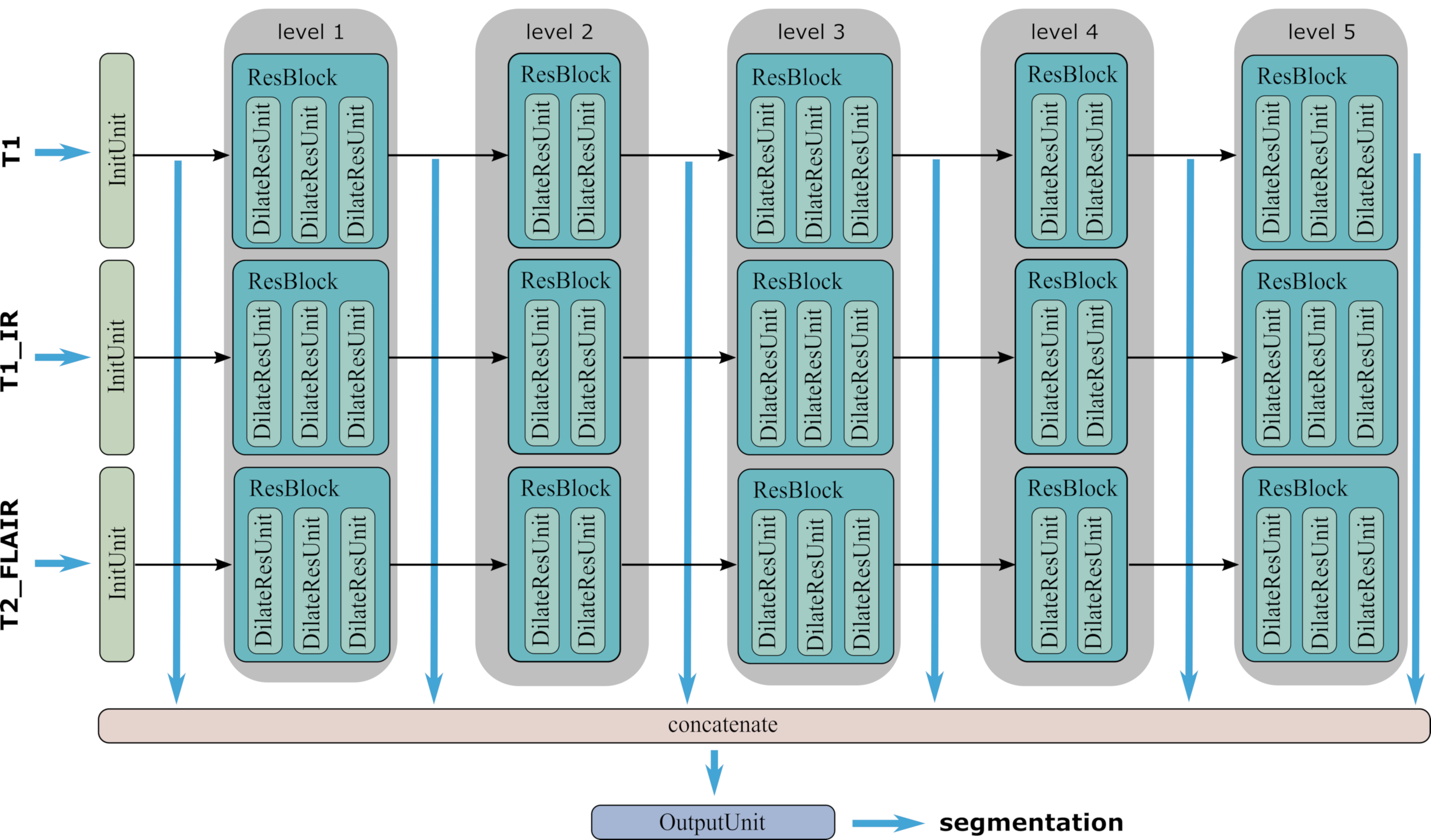}
		\caption{MixNetv3: modality streams are kept separate until the OutputUnit aggregates information from each stream.}
		\label{fig:v3}
	\end{minipage}
\end{figure}

The InitUnit consists of a single 5x5 convolutional layer and an optional pooling layer. Depending on the input channels, the convolution kernels can be of different sizes. In Figure \ref{fig:v1}, three modalities are stacked together, while mixNetv2 (Figure \ref{fig:v2}) and mixNetv3 (Figure \ref{fig:v3}) have three input streams. Thus, the kernel sizes are 5x5x3 and 5x5x1, respectively.  In addition, the pooling layer aims to reduce memory usage when necessary. If the pooling layer in the InitUnit is used, the upscaling layer in the OutputUnit should also be activated. In this work, we use a 2x2 pooling with stride 2.

\subsubsection{Residual Dilated Convolution Unit (DilateResUnit)}
\label{subsubsec:resUnit}

The training difficulty varies in different network architectures. For example, the degradation phenomenon arises in practice for a deeper plain CNN, although it includes the solution space of a shallower one. \cite{res1} conjecture that the deep plain CNN may have exponentially slow convergence rates and provides empirical evidence showing that a network composed of residual units is easier to optimize. The proceeding work \cite{res2} argues that the training procedure benefits from a "direct" path for information propagation, not only within a residual unit but through the whole network. Inspired by the successful works \cite{res1,res2}, we construct a deep residual learning network (DilateResUnit) with 'clear' paths through the layers and multiple modality streams for information propagation.

\begin{figure}
	\begin{minipage}{.33\textwidth}
		\centering
		\includegraphics[width=0.83\linewidth]{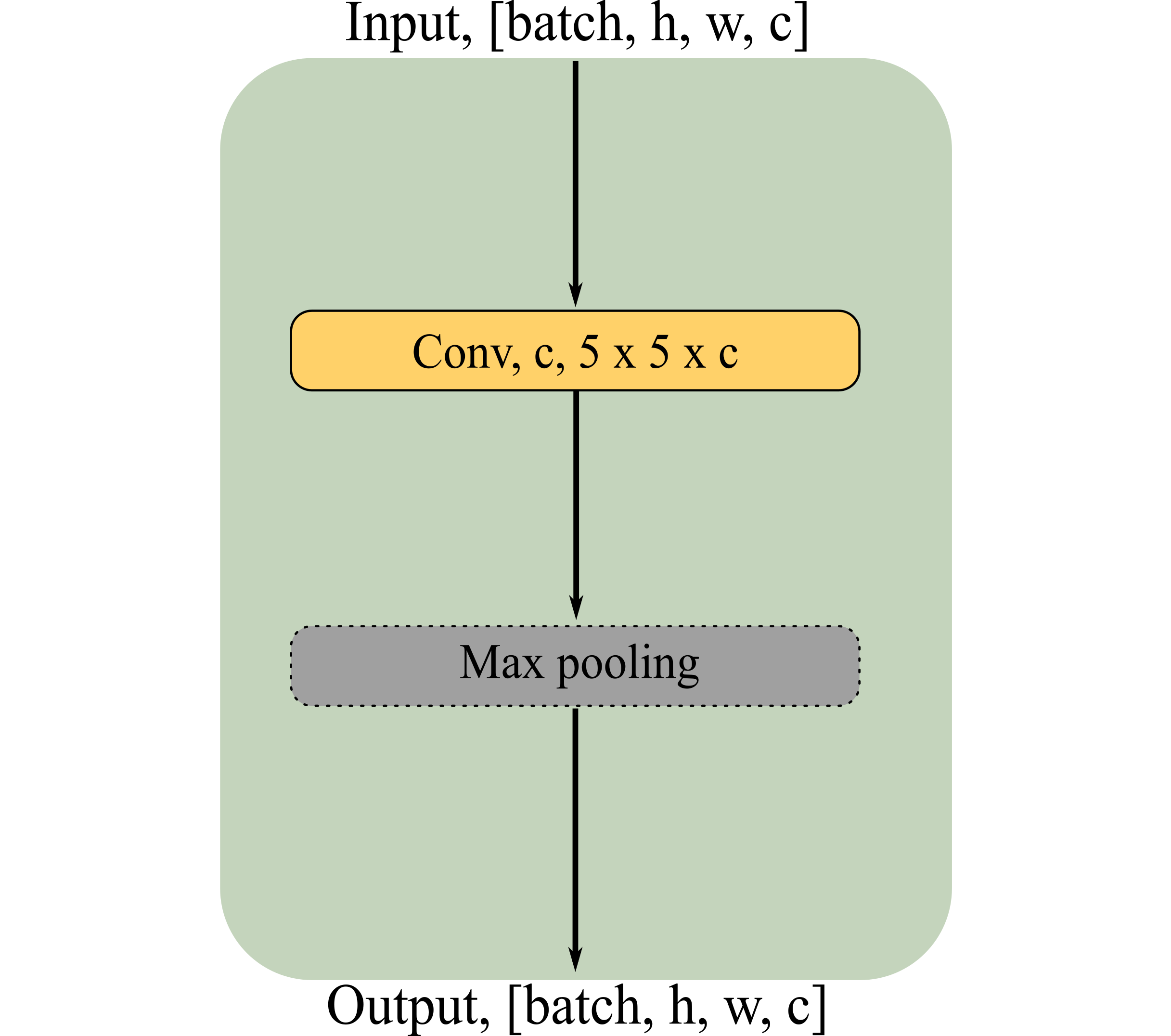}
		\caption{InitUnit}
		\label{fig:input}
	\end{minipage}%
	\begin{minipage}{.33\textwidth}
		\centering
		\includegraphics[width=0.88\linewidth]{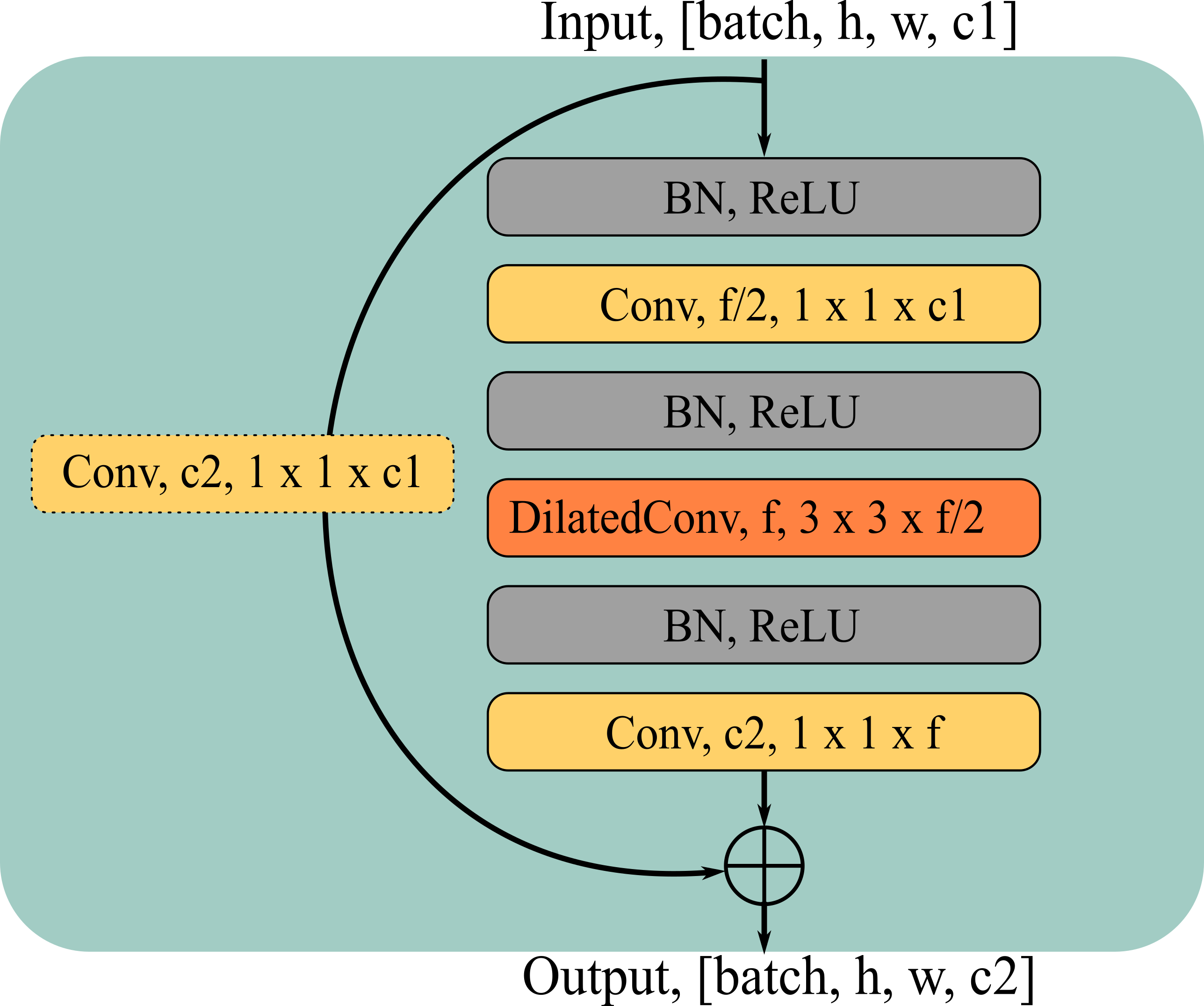}
		\caption{DilateResUnit}
		\label{fig:unit}
	\end{minipage}
	\begin{minipage}{.33\textwidth}
		\centering
		\includegraphics[width=0.838\linewidth]{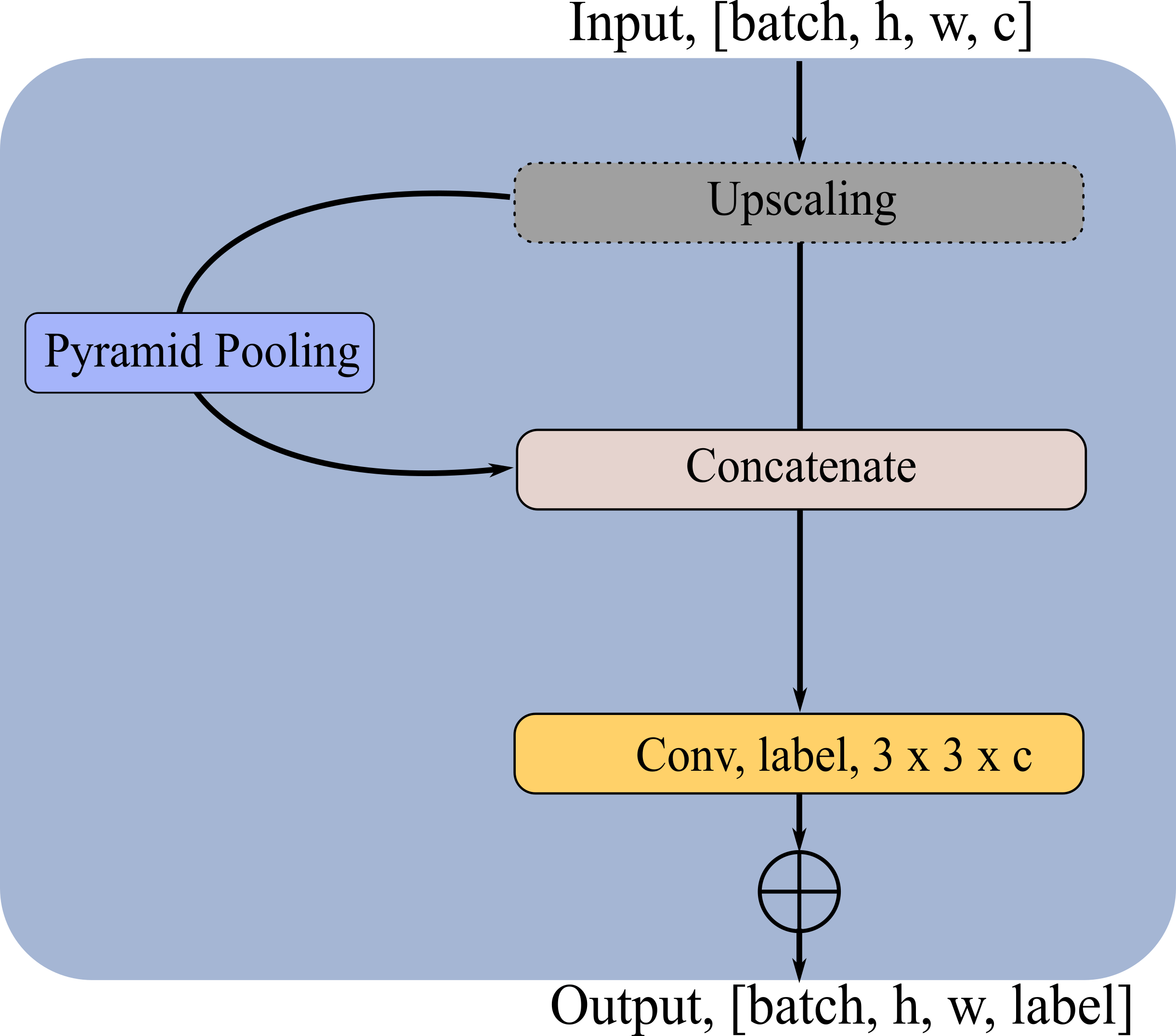}
		\caption{OutputUnit}
		\label{fig:output}
	\end{minipage}
\end{figure}

As shown in Figure \ref{fig:unit}, the shortcut lets the input feature map pass through the unit directly and only the differences between inputs and outputs are learned. When such units are connected to form a network, these short paths will also be interlinked throughout the network. Compared to the residual unit in \cite{res2}, the second convolutional layer is replaced by a dilated convolutional layer. Alternating convolutional layers and polling layers are a CNN common structure. The Pooling layer increases the receptive field efficiently while keeping the computational workload reasonable. However, the pooling layer loses localization information which is critical for segmentation tasks. Deconvolutions \cite{unet} and dilated convolutions (also known as atrous convolution) \cite{deepLab} are possible solutions. Different from the deconvolution where extra layers are involved to recover lost resolution, the dilated convolution keeps the resolution unchanged through the forward propagation. Extra layers mean more parameters. Assuming a network with less parameters is easier to train, we adopt the dilated convolution in this work.

A DilateResUnit is determined by four parameters: $c1, c2, f$ and $d$. The number of filters and the filter size of the dilated convolutional layer is $f/2$ and $f$, while $d$ is the dilation factor. The first and last 1x1 convolutional layers are determined by the channels of the input and output feature map. When the inputs and outputs are of different sizes, a 1x1 convolutional layer will be inserted on the shortcut. Since we use the same $f$ through the network, only the units before and after a concatenation in Figure \ref{fig:v2} (except the final concatenation) have such shortcut convolutions. 

\subsubsection{Output Unit (OutputUnit)}
\label{subsubsec:outputUnit}

As discussed in Section \ref{sec:intro}, anatomical structures preserve certain relative positions. Thus, the OutputUnit augments the input feature map with a global prior first, and then outputs results through a 3x3 convolutional layer.  The global prior is captured by a pyramid pooling module \cite{psp}. The pyramid pooling module separates the input feature map into sub-regions and forms representation by average pooling. Then, bilinear interpolation is performed to get the same size as the original feature map. In this work, we use a four-level pyramid with 2x2, 4x4, 6x6, 12x12 bins respectively.

Finally, the upscaling is performed to recover the original resolution, only when the pooling layer in the InitUnit is used. If the network can fit into the memory, pooling and upscaling are not necessary.

\subsection{Network Architecture}
\label{subsec:orga}

In this section, we discuss three styles of using multiple modalities: stacked channels, periodic summarization and parallel streams. Correspondingly, three network architectures (MixNetv1, MixNetv2 and MixNetv3) are constructed with the units introduced in Section \ref{subsec:unit} to test the effect on the results. 

At the output end, all three networks aggregate features form different levels. A multi-modality, mutli-scale feature map is then passed to the OutputUnit. which augments the feature map with a global prior and makes the final prediction. Detailed network parameters are listed in Table \ref{tab:parameters}.

To train the network, we compute the cross-entropy loss of each pixel in an image and accumulate them as the training loss. In this work, all pixels are treated equally, ingoring the label imbalance. The training process can run streadily in this way, but labels of a relatively small number may not receive enough attention. Weighing pixels of different labels is an approach worth trying.  

\subsubsection{Stacked Channels (MixNetv1)}
A straightforward way to fuse multiple modalities is to stack them as different channels. Thus, the input of MixNetv1 is a batch of 3-channel images. The forward propagation path is composed of serially connected DilateResUnits. Since the output of a DilateResUnit has a similar resolution with the input, we set the the filter number of all units to the same. In this way, the feature map size and the corresponding computation are balanced throughout different layers.
    
\subsubsection{Periodic Summarization (MixNetv2)}
MixNetv2 is a network architecture between MixNetv1 and MixNetv3. MixNetv1 fuses the multiple modalities at the very beginning, while MixNetv3 keeps different modality streams independent until the final output. In MixNetv2, periodic summarization of multi-modality information is performed. As shown in Figure \ref{fig:v2}, Level 1, Level 3 and Level 5 play such a role. The summarization is then fed back to each modality stream.

\subsubsection{Parallel Streams (MixNetv3)}
Three modality streams propagate forward independently in MixNetv3. Features from three streams are only collected when the OutputUnit makes the final prediction. Actually, the solution space of MixNetv3 is contained in MixNetv1. Each neuron in MixNetv1 has connection to all three modalities (indirect connections considered). If we force each neuron to connect to only one modality by setting some network parameters to 0, MixNetv1 can be equivalent to MixNetv3. However, MixNetv3 performs better than MixNetv1 based on our experiments. Experiment results are demonstrated in Section \ref{sec:evaluation}.     
  
\begin{table}
	\centering
	\caption{Parameters of three MixNet versions. The input channel $c1$, filter number $f$, dilation factor $d$ and output channel $c2$ of the DilateResUnit are listed with respect to the network level. As described in Section \ref{subsubsec:resUnit}, the DilatedResUnit is fully determined by these four parameters.}\label{tab:parameters}
	\begin{tabular}{*6c}
		\hline
		\textbf{MixNetv1} &  Level 1 & Level 2 & Level 3 & Level 4 & Level 5\\
		\hline
		Input &  120x120x72 & 120x120x72 & 120x120x72 & 120x120x72 & 120x120x72\\
		Filters &  72 & 72 & 72 & 72 & 72\\
		Dilation &  2 & 1 & 4 & 1 & 8\\
		Output &  120x120x72 & 120x120x72 & 120x120x72 & 120x120x72 & 120x120x72\\
		\hline
		\textbf{MixNetv2} &  Level 1 & Level 2 & Level 3 & Level 4 & Level 5\\
		\hline
		Input &  120x120x72 & 120x120x48 & 120x120x72 & 120x120x48 & 120x120x72\\
		Filters &  24 & 24 & 24 & 24 & 24\\
		Dilation &  2 & 1 & 4 & 1 & 8\\
		Output &  120x120x24 & 120x120x24 & 120x120x24 & 120x120x24 & 120x120x24\\
		\hline
		\textbf{MixNetv3} &  Level 1 & Level 2 & Level 3 & Level 4 & Level 5\\
		\hline
		Input &  120x120x24 & 120x120x24 & 120x120x24 & 120x120x24 & 120x120x24\\
		Filters &  24 & 24 & 24 & 24 & 24\\
		Dilation &  2 & 1 & 4 & 1 & 8\\
		Output &  120x120x24 & 120x120x24 & 120x120x24 & 120x120x24 & 120x120x24\\
		\hline
	\end{tabular}
\end{table}

\subsection{View MRI Volume from Different Directions}
\label{subsec:direction}

For a 2D CNN, the 3D MRI volume can be observed from any direction. The most commonly used are the three anatomical planes: the sagittal plane, the coronal plane and the transverse plane. By viewing the MRI volume from different directions, multiple batches of 2D slices can be acquired for training. For example, a 120x120x120 volume will generate 360 images of the three anatomical planes. In fact, more directions can be included.

On one hand, changing the observation direction provides more training images. On the other hand, fusing predictions is actually an ensemble model, which improves the algorithm robustness and benefit the spatial consistency.

The annotation resolution of the MRBrainS2018 dataset is anisotropic in three directions. Therefore, this strategy cannot be fully utilized. We train three networks on the sagittal, coronal, transverse plane and fuse the predictions. Further tests can be performed by training a single classifier with images acquired along different orientations.

\section{Results}
\label{sec:evaluation}
The experiments are performed with the MICCAI challenge MRBrainS2018 dataset. The challenge releases 7 MRI scans (including T1-weighted, T1-weighted inversion recovery and T2-FLAIR) as the training data. Another 23 scans are kept secret for test. We test the three networks using leave-one-out cross validation strategy with the training dataset. MixNetv2 is submitted to the challenge and an evaluation of MixNetv2 on the test dataset is performed by the challenge organizers.
\subsection{Preprocessing and Data Augmentation}
Bias field correction \cite{bias} and image registration are performed by the challenge organizer. In addition to this, we linearly scale each modality image of each scan to have zero mean and unit variance.

To train the very deep network, the data is heavily augmented with elastic deformation \cite{elas}, scaling, rotation and translation. As for the sagittal and coronal plane, the resolution in horizontal and vertical directions are four times different. Thus, we only apply flipping and translation.

It is worth mention that excessive elastic deformation and scaling may lead to an unstable training. We use scaling factors of 0.9, 0.95, 1.05 and 1.1, elastic deformation factor $\alpha=10$ and $\sigma=4$ \cite{elas} in this work. Rotation is performed around the image center with 8 degrees: \ang{0}, \ang{45}, \ang{90}, \ang{135}, \ang{180}, \ang{225}, \ang{270} and \ang{315}. The random translation is limited to 0.15 percent of the image size. We use all augmentation methods separately, that is, no images are generated from augmented images.

\subsection{Training}

The network is trained with gradient descent optimization algorithm with Nesterov momentum. The momentum is set to 0.99. The initial learning rate is 2e-4 and is halved after each preset boundary epoch, which is 0.2, 0.4, 0.6, 0.75, 0.8, 0.85, 0.9 and 0.95 of the total number of training epochs. L2 regularization is used to prevent overfitting with a weight decay of 1e-3.

\subsection{Evaluation and Conclusion}
The results are evaluated according to three metrics: Dice coefficient (Dice), 95th-percentile Hausdorff distance (HS) and Volumetric similarity (VS). Additionally, a sum of weighted metrics is computed as the overall score for MRBrainS ranking. Details of the evaluation metrics and the overall score are described in \cite{mrbrains2018}.  

\begin{figure}
	\centering
	\begin{minipage}[b]{0.46\textwidth}
		\includegraphics[width=\linewidth]{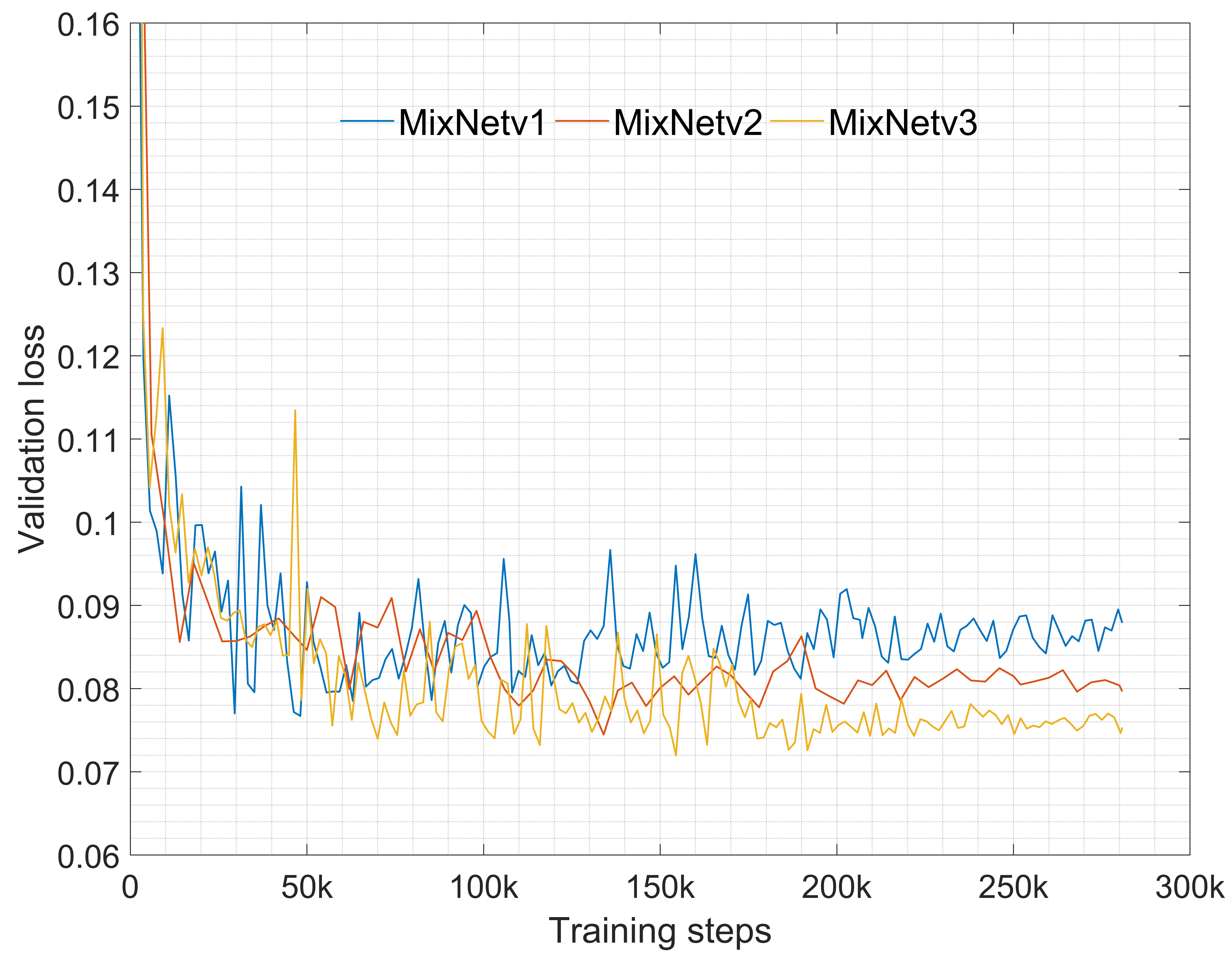}
		\caption{Validation loss during training (subject 1 as the validation data).}
		\label{fig:val}
	\end{minipage}
	\hfill
	\begin{minipage}[b]{0.46\textwidth}
		\includegraphics[width=\linewidth]{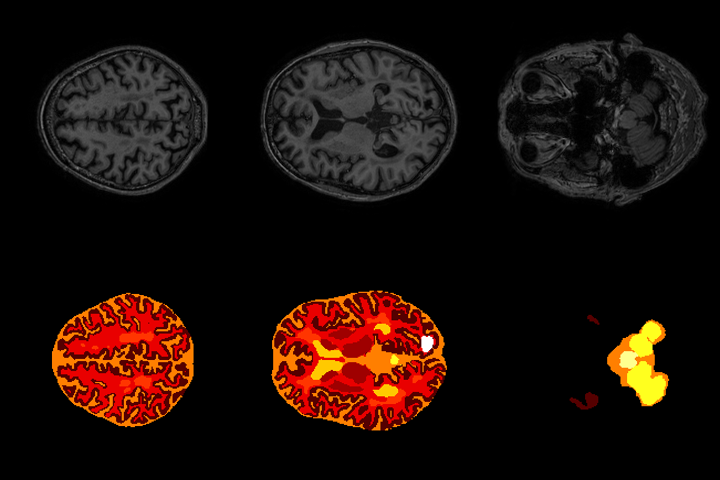}
		\vspace{0.1cm}
		\caption{Qualitative segmentation results of 8 brain structures.}
		\label{fig:res}
	\end{minipage}
\end{figure}

To compare the performance of three network variants, we run the leave-one-out cross validation as a 3-label segmentation problem (GM, WM and CSF) on the MRBrainS2018 training dataset. As shown in Table \ref{tab:res_val}, MixNetv3 gives the best results. The cross validation results of MixNetv1 and MixNetv2 are quite close. But MixNetv2 has a lower validation loss (see Figure \ref{fig:val}). As discussed in \ref{subsec:orga}, MixNetv1 contains the solution space of MixNetv3. However, the results of MixNetv1 is worse. We conjecture that the architecture of parallel modality streams can learn complementary features more easily.

By MixNetv2\_multi, three classifiers are trained on the sagittal plane, the coronal plane and the transverse plane, respectively. Results are obtained by fusing predictions of three MixNetv2 classifiers with the corresponding weights 1:1:4. The weights are empirically chosen based on the fact that the transverse plane resolution is 4 times higher. Although the classifiers on the sagittal plane and the coronal plane performs much worse, the fused results still improves.

\begin{figure}
	\begin{minipage}{\textwidth}
		\centering
		\includegraphics[width=\linewidth]{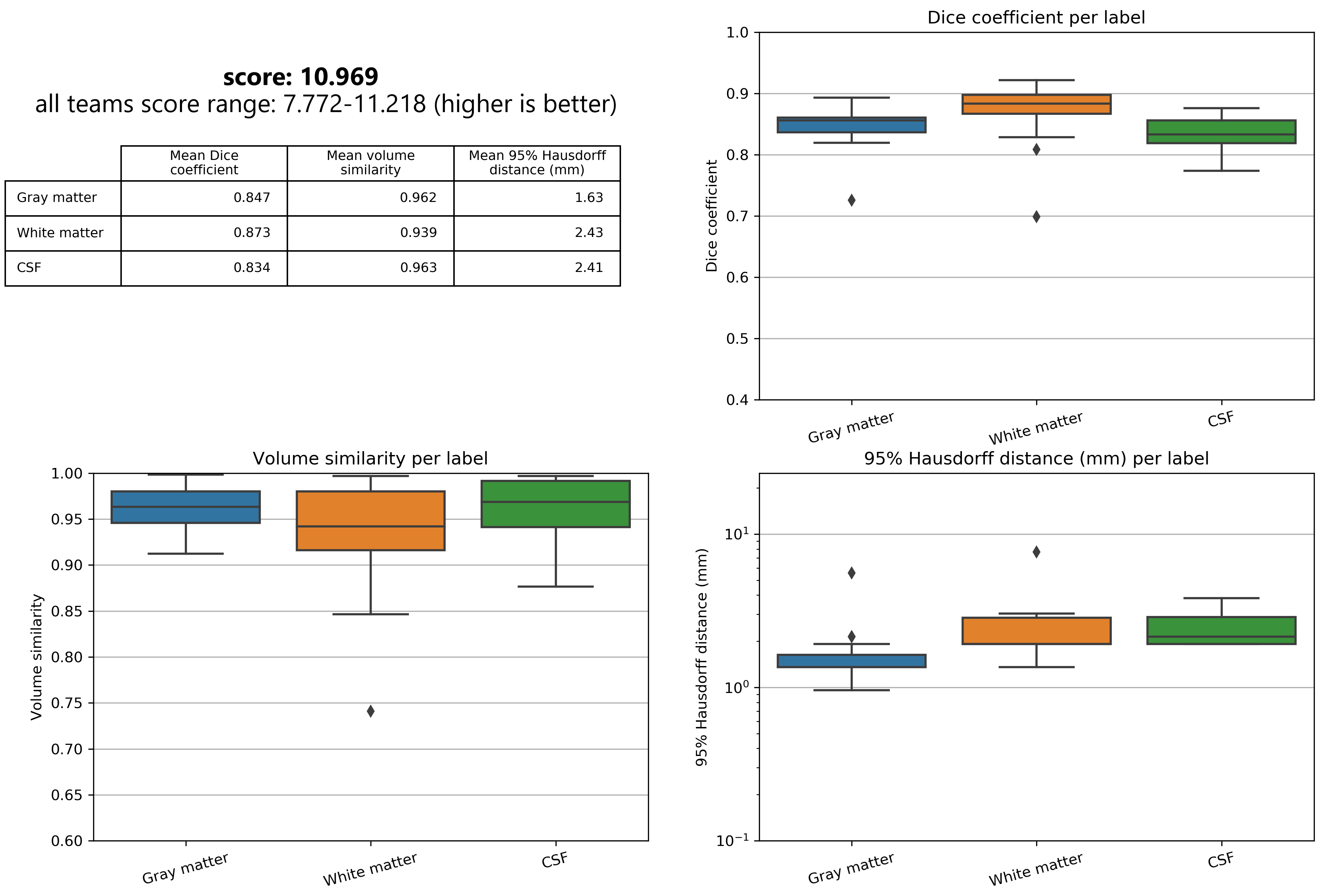}
		\caption{Test results of MixNetv2\_multi on MRBrainS2018 test dataset (3-label task).}\label{fig:res_test3}
	\end{minipage}
\end{figure}

\begin{figure}
	\begin{minipage}{\textwidth}
		\centering
		\includegraphics[width=\linewidth]{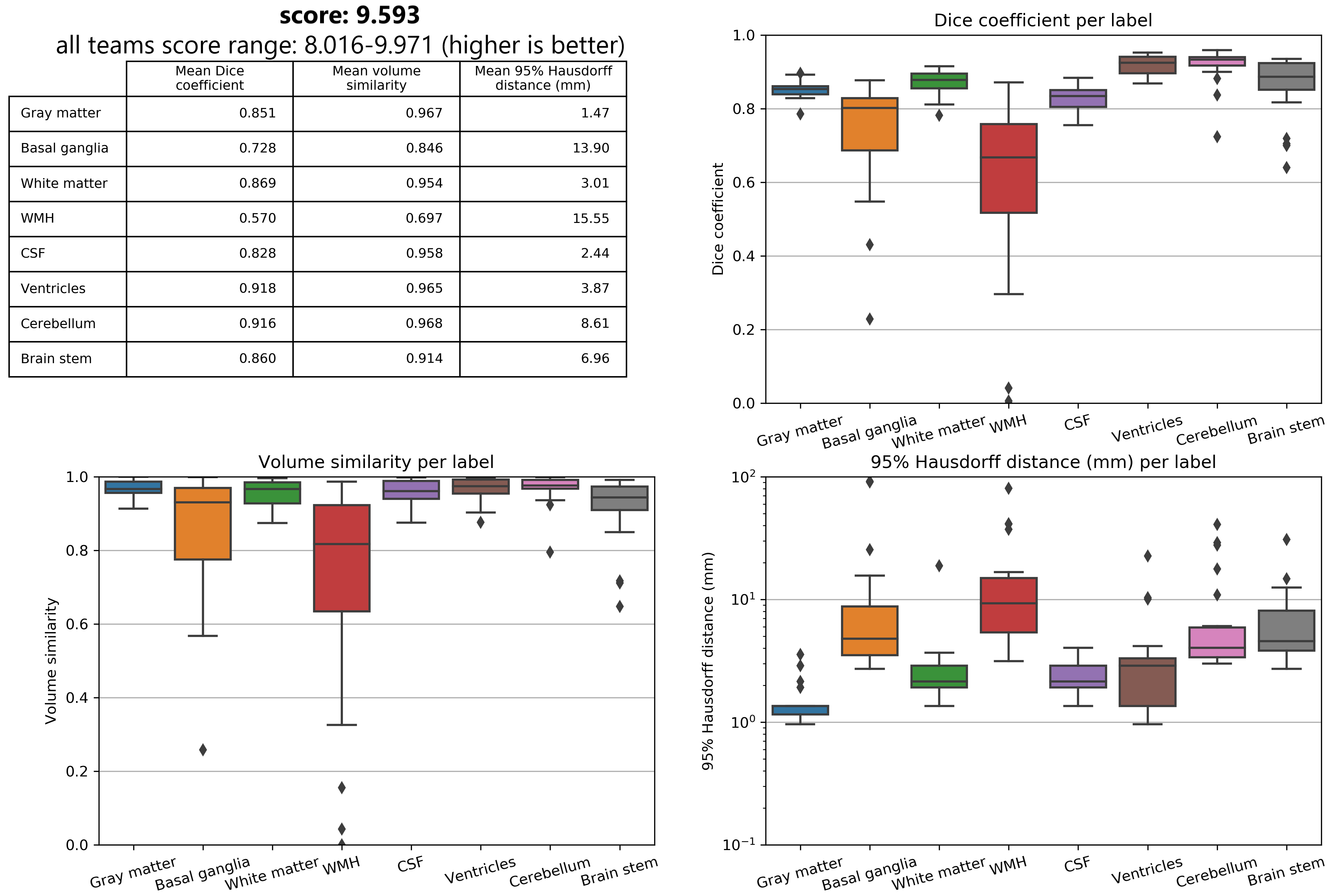}
		\caption{Test results of MixNetv2\_multi on MRBrainS2018 test dataset (8-label task).}\label{fig:res_test8}
	\end{minipage}
\end{figure}

MixNetv2\_multi was also trained with the full training dataset as a 3-label and 8-label task. Figure \ref{fig:res} shows the qualitative results of 8-label predictions by MixNetv2\_multi. Trained models were submitted to the challenge. Figure \ref{fig:res_test3} and Figure \ref{fig:res_test8} show the evaluation performed by the challenge organizer on the test dataset. We notice a performance drop between the validation results and the evaluation results (about 0.02). That is reasonable, because the relatively small training dataset may not cover all the cases very well.  

In the 8-label segmentation task, our network has difficulties in classifying WHM and basal ganglia. One possible reason is the imbalance of labels in the training data. We do not use any methods of balancing the labels during training, that is, labels with a small number in the training data are easy to ignore. The 8-label methods taking part in the MRBrainS2018 challenge differ mainly in the performance of segmenting WHM and basal ganglia. This problem deserves further study.  

\begin{table}
	\centering
	\caption{Cross validation results of MixNetv1, MixNetv2 and MixNetv3, performed on the MRBrainS2018 training dataset. The network is trained as a 3-label segmentation task (WM, GM and CSF).}\label{tab:res_val}
	\begin{tabular}{*{10}{c}}
		\hline
		{} & \multicolumn{3}{c}{GM} & \multicolumn{3}{c}{WM} & \multicolumn{3}{c}{CSF} \\
		{} &  Dice & HD & VS &  Dice & HD & VS &  Dice & HD & VS \\
		\hline
		\textbf{MixNetv1} &  .8524 & .9583 & .9728 &  .9000 & 1.9167 & .9759 &  .8599 & 1.9167 & .9508 \\
		\textbf{MixNetv2} &  .8500 & .9583 & .9772 &  .8966 & 1.9167 & .9626 &  .8609 & 1.9167 & .9506 \\
		\textbf{MixNetv2\_multi} &  .8511 & .9583 & .9762 &  .9001 & 1.3553 & .9689 &  .8624 & 1.9167 & .9447 \\
		\textbf{MixNetv3} &  .8557 & 0.9583 & .9789 &  .9049 & 1.3552 & .9743 &  .8609 & 1.9167 & .9578 \\
		\hline
	\end{tabular}
\end{table}

\section{Summary}
In this work, we propose the MixNet, a deep residual CNN to tackle the brain structure segmentation problem. The network achieves state-of-the-art results with a relatively small training dataset. Three variants of MixNet is tested to see the effect of different modality mixing styles. Based on the experiment results, the network of parallel modality streams shows better performance, which implies that learning complementary features may be easier for this architecture. 

As future work, a single classifier trained with images acquired along different orientations of the 3D MRI volume is worth testing. To do this, either a dataset of isotropic annotation resolutions is available or the resolution difference is tackled properly.

%
%
%
%

\end{document}